\documentclass[aps,showpacs,superscriptadress,preprint]{revtex4}
\usepackage{graphicx}

\begin{document}
\title{Decay of massive scalar field in a black hole background immersed in  magnetic field}

\author{Chen Wu$^{1}$\footnote{Electronic address: wuchenoffd@gmail.com}} \affiliation{
\small 1. Shanghai Institute of Applied Physics, Chinese Academy of
Sciences, Shanghai 201800, China}
\author{\footnotesize Renli Xu$^{2}$\footnote{Electronic address: xurenli.phy@gmail.com}}
\affiliation{ \small 2. Key Laboratory of Modern Acoustics and Department of Physics, Nanjing University, Nanjing 210093, China\\}

\begin{abstract}
We evaluate   quasinormal modes of massive scalar field of the Ernst spacetime, an exact solution of the Einstein-Maxwell equations describing a black hole immersed in
uniform magnetic field $B$.  It is known that the quasinormal spectrum for massive scalar field in the vicinity of the magnetized black holes acquires an
effective mass $\mu_{eff}= \sqrt{4B^2 m^2+\mu^2}$, where $m$ is the azimuthal number and $\mu$ is the mass of scalar field. The numerical result shows that increasing of the field effective mass and the magnetic field $B$ gives rise to decreasing of the imaginary part of the quasinormal modes until reaching the vanishing  damping rate.
\end{abstract}

 \maketitle

\section{Introduction}
It has been well understood that when  a classical black hole is perturbed by an exterior field, the dynamical evolution of the field will undergo
 three stages \cite{black hole}. The first one is an initial wave burst coming directly from the source and is dependent  on the initial form of the original  wave  field. The second one involves the damped oscillations called the quasinormal modes (QNMs), which do not depend on the initial values of the wave but are characteristic of the background black hole spacetimes. The QNMs are defined as the complex solutions to the perturbation wave equations under certain boundary conditions. The QNM frequencies have complex values because of radiation damping. The last stage is the powerlaw tail behavior.
  Here we would like to concentrate on the intermediate stage of the evolution of the massive scalar field where the QNMs dominate.

Astrophysical interests in QNMs originated from their relevance in gravitational wave analysis.
 The QNMs of black holes are expected to be detected by gravitational wave detectors such as LISA \cite{LISA}.
  Recently the study of the QNMs has gained considerable attention
  coming from AdS/CFT correspondence in string theory . The lowest quasinormal frequencies of black holes have a direct interpretation as dispersion relations of hydrodynamic excitations in the ultra-relativistic heavy ion collisions \cite{hydro}. All of these aspects motivated the  extensive numerical and analytical study of QNMs for different spacetime and different fields (both massive and massless) around black holes. We refer the reader to the reviews \cite{review} where a lot of references to the recent research of QNMs can be found.

It is well known that  that the massive QNMs decay  more slowly than the massless one in a lots of references, both by frequency domain method \cite{will,Konoplya} and by time domain method\cite{Koyama, Wang}.
In Ref. \cite{will}, Simone and Will have investigated massive QNMs on Schwarzschild black hole spacetime using the WKB method \cite{WKB}.
They studied the dependence of QNM frequencies on the mass of the scalar field, which is restricted to a narrow range due to the restriction required by the WKB
method. And a fully revelation of the dependence of QNM frequencies on a wide range of field mass is still expected.
 Ohashi and Sakagami \cite{resonance} investigated QNMs for the decay of the massive scalar field on the Reissner-Nordstr\"{o}m black hole spacetime by using the continued fraction method \cite{Leaver} and found that
there are QNMs with arbitrary long life when the field mass has special values. They named these modes as quasi-resonances modes (QRMs).

The magnetic field is one of the most important constituents of the cosmic space and one of the main sources of the dynamics of interacting matter in the universe.
It was found that the equation of state of compact stars is strongly affected by the strong magnetic field \cite{wuchen}. Moreover,
strong magnetic fields of up to $10^4-10^8G$ are supposed to exist near supermassive black holes in the active galactic nuclei and even around stellar
mass black holes \cite{SNZhang}.  Interaction between a black hole and magnetic field can happen in a lot of physical situations, such as an charged accretion disk or other charged matter distributions around black hole. If mini-black holes are created in ultra-relativistic particle collisions in the brane-world scenarios, it exists possibility of interaction between strong magnetic fields and mini-black holes in a great variety of high energy processes when quantum gravity states are excited.\
So astrophysicists are highly interested in investigating the magnetic fields around black holes. In Ref. \cite{Magnetized BH QNM}, Konoplya and his coworkers have found the QNMs for the Ernst black hole \cite{Ernst} that is a black hole immersed in  an external magnetic field. They described the influence of the  magnetic field onto characteristic quasinormal spectrum of black holes.

In this paper, we consider the massive scalar field perturbations around the Ernst black hole.  We shall find the QNMs of the Ernst black holes in frequency domain. Our numerical investigation shows that the magnetic field $B$ increases with the decrement of the imaginary part of the QNM until reaching the vanishing damping rate.
When some threshold values of $B$ are exceeded, the particular QNMs disappear. In sect. 2, we consider the Klein-Gordon equation in the Ernst spacetime and its reduction into Schrodinger-like equation with a particular effective potential. Then we evaluate the quasinormal frequencies of the massive scalar field using the continued fraction method.  The last section ends up this paper with summary and conclusion.

\section{The basic equations and numerical results }
In 1976 Ernst found a class of exact black hole solutions of the
Einstein¨CMaxwell equations \cite{Ernst}. The simplest of these solutions corresponds
to a magnetized Schwarzschild black hole, also known as the
Ernst metric, which has the form

\begin{eqnarray}
ds^2 = \Lambda^2 \left(  -\left(1-\frac{2M}{r}\right)dt^2  + \left( 1-\frac{2M}{r}\right)^{-1}dr^2  + r^2d\theta^2 \right) + \frac{r^2\text{sin}^2\theta}{\Lambda^2}d\phi^2 ,
\end{eqnarray}
where the external magnetic field $B$ is determined by the relation
\begin{eqnarray}
\Lambda^2 =  1+ B^2r^2\textmd{sin}^2\theta.
\end{eqnarray}
The vector potential giving rise to the homogeneous magnetic field reads:
\begin{eqnarray}
A_\mu dx^\mu = - \frac{Br^2 \textmd{sin}^2\theta}{\Lambda}d\phi.
\end{eqnarray}
As a magnetic field is assumed to exist everywhere in space, the above metric is not asymptotically flat.

The Klein-Gordon equation describing the evolution of massive scalar perturbation field outside Ernst black hole is given by
\begin{eqnarray}
   \frac{1}{\sqrt{-g}} \partial_\mu(g^{\mu\nu}\sqrt{-g} \partial_\nu\Phi)
 + \mu^2 \Phi = 0.
\end{eqnarray}
Generally, the Klein-Gordon equation  for the Ernst black hole spacetime does not allow separation of radial and angular variables. Yet because of  small $B$
in our problem,  one can safely neglect terms higher than $B^2$ in Eq. (4). In this way it is known that only dominant correction due to the magnetic field to the effective potential of the Schwarzschild black hole approximately  \cite{Magnetized BH QNM}. The  Klein-Gordon equation for the angular part has the form

\begin{eqnarray}
\frac{P_{sch}(\theta, \phi)\Phi}{r^2} + \frac{\Lambda^4-1}{r^2\textmd{sin}^2\theta} \partial_{\phi\phi} \Phi = 0 ,
\end{eqnarray}
with $P_{sch}(\theta, \phi)$ meaning the corresponding pure-Schwarzschild part of the angular equation, $P_{sch}(\theta, \phi) \Phi = -l(l+1)\Phi$, then one find
\begin{eqnarray}
\frac{1}{\textmd{sin}\theta}\frac{\partial}{\partial\theta} \left(\textmd{sin}\theta \frac{\partial Y_{lm}}{\partial \theta} \right) +
\frac{1}{\textmd{sin}^2\theta}  \frac{\partial^2 Y_{lm}}{\partial^2 \phi} = \left(-l(l+1) + 4B^2m^2 r^2 \right)Y_{lm}.
\end{eqnarray}

After separation of the angular variables, one can reduce the wave equation  (4) in the Ernst background to the Schr\"{o}dinger wave equation
\begin{eqnarray}
\left( \frac{d^2}{dr^{*2}} + \omega^2 - V(r^*) \right) \Psi(r^*) = 0,
\end{eqnarray}
with the effective potential $V(r)$:
\begin{eqnarray}
V(r) = f(r) \left(  \frac{l(l+1)}{r^2} + \frac{2M}{r^3} + 4B^2m^2 + \mu^2 \right),
\end{eqnarray}
where
\begin{eqnarray}
f(r)  = 1- \frac{2M}{r},   \,\,\,\,\,    dr^* = \frac{dr}{f(r)},
\end{eqnarray}
and $m$ is the azimuthal quantum number. One can see that the effective potential Eq. (8) coincides with the potential for the massive scalar field with the effective mass $\mu_{eff} = \sqrt{4B^2m^2 + \mu^2}$ in the Schwarzschild background.

The wave equation with the obtained potential Eq. (8) has  the sub-dominant asymptotic term at infinity:
\begin{eqnarray}
\psi (r^*) \sim C_+ e^{i\chi r^*}r^{iM \mu_{eff}^2/\chi} (r, r^* \rightarrow +\infty),
\end{eqnarray}
\begin{eqnarray}
  \chi = \sqrt{\omega^2 - \mu_{eff}^2}
\end{eqnarray}
 Within the continued fraction method we can calculate the singular factor of the solution of Eq. (7) that satisfies in-going wave boundary condition at the horizon and  Eq. (10) at infinity, and  expand the remaining part into the Frobenius series that
are convergent in the R-region ($ -\infty< r^* <+\infty$). The solution of Eq. (7) is expanded as follows:
\begin{eqnarray}
\psi(r) = e^{i\chi r} r^{(2iM\chi + iM\mu_{eff}^2/\chi)} \left( 1-\frac{2M}{r} \right)^{-2iM\omega} \sum_n a_n \left( 1 - \frac{2M}{r} \right)^n,
\end{eqnarray}

Substituting Eq. (12) into Eq. (7) we obtain the following recursion relation for the coefficients $a_n$:
\begin{eqnarray}
\alpha_0 a_1 + \beta_0 a_0 = 0, \,\,\,\,\,\,\, \alpha_n a_{n+1} + \beta_n a_n + \gamma_n a_{n-1} = 0, \,\,\,\, n>0,
\end{eqnarray}
where
\begin{eqnarray}
\alpha_n = (n+1)(n+1-4M\omega i),
\end{eqnarray}
\begin{eqnarray}
 \beta_n = -2n (n+1) - 1 - l(l+1) + \frac{M(\omega+\chi)(4M(\omega+\chi)^2  + i (2n+1)( \omega+3\chi) )}{\chi},
\end{eqnarray}
\begin{eqnarray}
\gamma_n = (n - Mi(\omega+\chi)^2/\chi)^2
\end{eqnarray}

Since the series are convergent at infinity, the ratio of successive $a_n$ will be given by the infinite continued fraction:
\begin{eqnarray}
\frac{a_{n+1}}{a_n} = \frac{\gamma_n}{\alpha_n} \frac{\alpha_{n-1}}{\beta_{n-1}- \frac{\alpha_{n-2}\gamma_{n-1}}{ \beta_{n-2}- \alpha_{n-3}\gamma_{n-2}/\cdots }}
- \frac{\beta_n}{\alpha_n} = - \frac{\gamma_{n+1}}{\beta_{n+1}- \frac{\alpha_{n+1}\gamma_{n+2}}{ \beta_{n+2}- \alpha_{n+2}\gamma_{n+3}/\cdots }}.
\end{eqnarray}

Thus QNM frequencies are given by the vanishing point of the following continued fraction equation:
\begin{eqnarray}
\beta_n - \frac{\alpha_{n-1} \gamma_n} {\beta_{n-1}- \frac{\alpha_{n-2}\gamma_{n-1}}{ \beta_{n-2}- \alpha_{n-3}\gamma_{n-2}/\cdots }}
 = - \frac{\gamma_{n+1} \alpha_n}{\beta_{n+1}- \frac{\alpha_{n+1}\gamma_{n+2}}{ \beta_{n+2}- \alpha_{n+2}\gamma_{n+3}/\cdots }}.
\end{eqnarray}

The quasinormal modes for massive scalar fields were studied for the first time by Will and Simone \cite{will} and late in \cite{Konoplya} with the help of the WKB method \cite{WKB}. The massive QNMs are characterized by the growing of the damping time with the mass until the appearance of the infinitely long lived modes called quasi-resonances \cite{resonance}.

\begin{table}[tbh]\centering
\caption{Fundamental quasinormal modes for Ernst black holes for
different values of the magnetic field $B$ . Here, $M=1,\mu=0$.
\vspace{0.3cm}} \label{table1}
\begin{tabular*}{15.0cm}{*{4}{c @{\extracolsep\fill}}}
\hline\hline
$B$  &$l=1, m=1$  &$l=2,m=1$ &$l=2,m=2$ \\
\hline
$0.005$  &0.292981 -- 0.097633i  &0.483675 -- 0.096748i &0.483770 -- 0.096716i  \\
$0.025$  &0.294054 -- 0.096988i  &0.484433 -- 0.096488i &0.486804 -- 0.095675i  \\
$0.050$  &0.297416 -- 0.094957i  &0.486804 -- 0.095675i  &0.496327 -- 0.092389i \\

$0.075$  &0.303040 -- 0.091521i  &0.490764 -- 0.094312i  &0.512346 -- 0.086795i \\
$0.100$  &0.310957 -- 0.086593i  &0.496327 -- 0.092389i &0.535100 -- 0.078676i  \\
$0.125$  &0.321199 -- 0.080040i  &0.503512 -- 0.089891i  &0.564937 -- 0.067625i \\
\hline\hline
\end{tabular*}
\end{table}

\begin{table}[tbh]\centering
\caption{Fundamental quasinormal modes for Ernst black holes for
different values of the scalar field mass for given $B$. The
parameter $ m$ is set to be $1$. \vspace{0.3cm}} \label{table2}
\begin{tabular*}{12.0cm}{*{4}{c @{\extracolsep\fill}}}
\hline\hline $l$ &$B$  &$\mu$ &$\omega$\\
\hline
1  &0.1  &0 & 0.310957 -- 0.086593i  \\
  &  &0.1 & 0.315507 -- 0.083712i  \\
 &  &0.2 & 0.329199 -- 0.074756i  \\
  &  &0.3 & 0.352055 -- 0.058663i  \\
 &  &0.4 & 0.383561 -- 0.033446i  \\
&  &0.5 & 0.415513 -- 0.010131i  \\
\hline
1  &0.2  &0 & 0.365601 -- 0.048286i  \\
\hline\hline
\end{tabular*}
\end{table}

\begin{table}[tbh]\centering
\renewcommand{\arraystretch}{0.9}
\begin{tabular*}{12.0cm}{*{4}{c @{\extracolsep\fill}}}
\multicolumn{4}{c}{\footnotesize TABLE \uppercase\expandafter{\romannumeral2.}~~~(continued.)}\\
\hline\hline $l$ &$B$  &$\mu$ &$\omega$\\
\hline
  &  &0.1 & 0.370064 -- 0.044731i  \\
 &  &0.2 & 0.383561 -- 0.033446i  \\
  &  &0.3 & 0.406864 -- 0.018802i  \\
 &  &0.4 & 0.431849 -- 0.006525i  \\
\hline
2  &0.1  &0 & 0.496327 -- 0.092389i  \\
  &  &0.1 & 0.499516 -- 0.091283i  \\
 &  &0.2 & 0.509127 -- 0.087926i  \\
  &  &0.3 & 0.525301 -- 0.082200i  \\
 &  &0.4 & 0.548279 -- 0.073859i  \\
  &  &0.5 & 0.578409 -- 0.062441i  \\
    &  &0.6 & 0.616040 -- 0.047068i  \\
 &  &0.7 & 0.660590 -- 0.026323i  \\
  &  &0.8 & 0.712166 -- 0.008647i  \\
\hline
2  &0.2  &0 & 0.535100 -- 0.078676i  \\
  &  &0.1 & 0.538382 -- 0.077485i  \\
 &  &0.2 & 0.548279 -- 0.073859i  \\
  &  &0.3 & 0.564937 -- 0.067625i  \\
 &  &0.4 & 0.588593 -- 0.058421i  \\
  &  &0.5 & 0.619495 -- 0.045577i  \\
    &  &0.6 & 0.657356 -- 0.027823i  \\
&  &0.7 & 0.696345 -- 0.010999i  \\
\hline
2  &0.3  &0 & 0.602270 -- 0.052867i  \\
  &  &0.1 & 0.605705 -- 0.051442i  \\
 &  &0.2 & 0.616040 -- 0.047068i  \\
  &  &0.3 & 0.633343 -- 0.039449i  \\
 &  &0.4 & 0.657356 -- 0.027823i  \\
  &  &0.5 & 0.695061 -- 0.010504i  \\
\hline\hline
\end{tabular*}
\end{table}
We obtained the QNMs for different values of $B, l $ and $m$. The results are summarized in Table 1. From Table 1 one can see that when the magnetic field $B$ increases,  the real part of $\omega$ grows and the imaginary part of  $\omega$ decreases. Therefore  black hole immersed in a strong magnetic field  is a better oscillator, i.e., in such case the quality factor $Q \sim \textmd{Re} \, \omega/ \textmd{Im} \, \omega$  increases, compared to the nonmagnetic circumstance. Fig. 1 shows that increasing of the the magnetic field $B$ gives rise to decreasing of the imaginary part of the QNM until reaching the vanishing damping rate. When some threshold values of $B$ are exceeded, the particular QNMs disappear.

Then for given  magnetic filed, We calculated the QNMs for Ernst black holes for different values of the scalar field mass in the two cases of $l =1 $ and 2.  We listed our numerical results in Table 2. The numerical investigation shows that increasing of the field mass $\mu$ also gives rise to decreasing of the imaginary part of the QNM until reaching the vanishing damping rate. When some threshold values of $\mu$ are exceeded, the quasinormal modes also disappear. For a given $l$ , for example $l=1$ and 2, the  bigger magnetic field $B $ is, the  smaller the threshold value of $\mu$ become. In order to illustrate our results more transparently, we show our numerical results in  Fig. 2.

\section{summary}
In Summary, we have presented in this paper the quasinormal modes of
massive scalar field of the Ernst black holes. Calculations show
that the real oscillation frequency grows when the magnetic field
increases. The damping rate decreases with the increment of the
magnetic field, so that the magnetized black hole is characterized
by longer lived modes with higher oscillation frequencies, i.e., by
larger quality factor. It is shown that the effective mass
$\mu_{eff} = \sqrt{4B^2m^2+ \mu^2}$ of the scalar field has crucial
influence on damping rate of the QNMs. In particular, the greater
the effective mass of the field is, the less the damping rate
becomes. As a result, purely real modes which corresponds to
non-damping oscillations appear, and when the field effective mass
is greater than certain threshold value, the corresponding QNMs
disappear. In Ref. \cite{Brito}, Brito and his cooperators have done a fully consistent study of massless
scalar field perturbations of Ernst black holes (including the
rotating version of these solutions), without performing the small-$B$
approximation. Next we will research their fully-consistent linear analysis and consider  how  their method influence the behavior of the QRMs.

\begin{acknowledgments}
We thank Prof. Ru-Keng Su  for very helpful discussions.
This work is supported partially by the Major State
Basic Research Development Program in China (No. 2014CB845402).
\end{acknowledgments}

\begin{figure}[tbp]
\includegraphics[width=13cm,height=19cm]{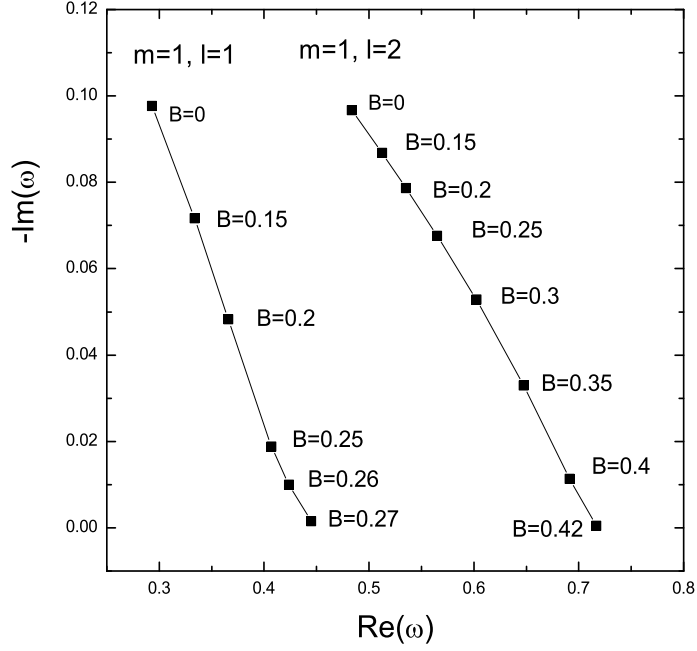}
\caption{ Fundamental quasinormal modes of the massless scalar field for Ernst black holes  under different magnetic field $B$, for $m =1, l = 1, 2$.
    Given $l = 1$  for example, the  bigger magnetic field $B$ is, the smaller the imaginary part of QNM becomes. When the threshold value of $B = 0.27$ is exceeded, the particular QNMs disappear. }
\end{figure}

\begin{figure}[tbp]
\includegraphics[width=13cm,height=19cm]{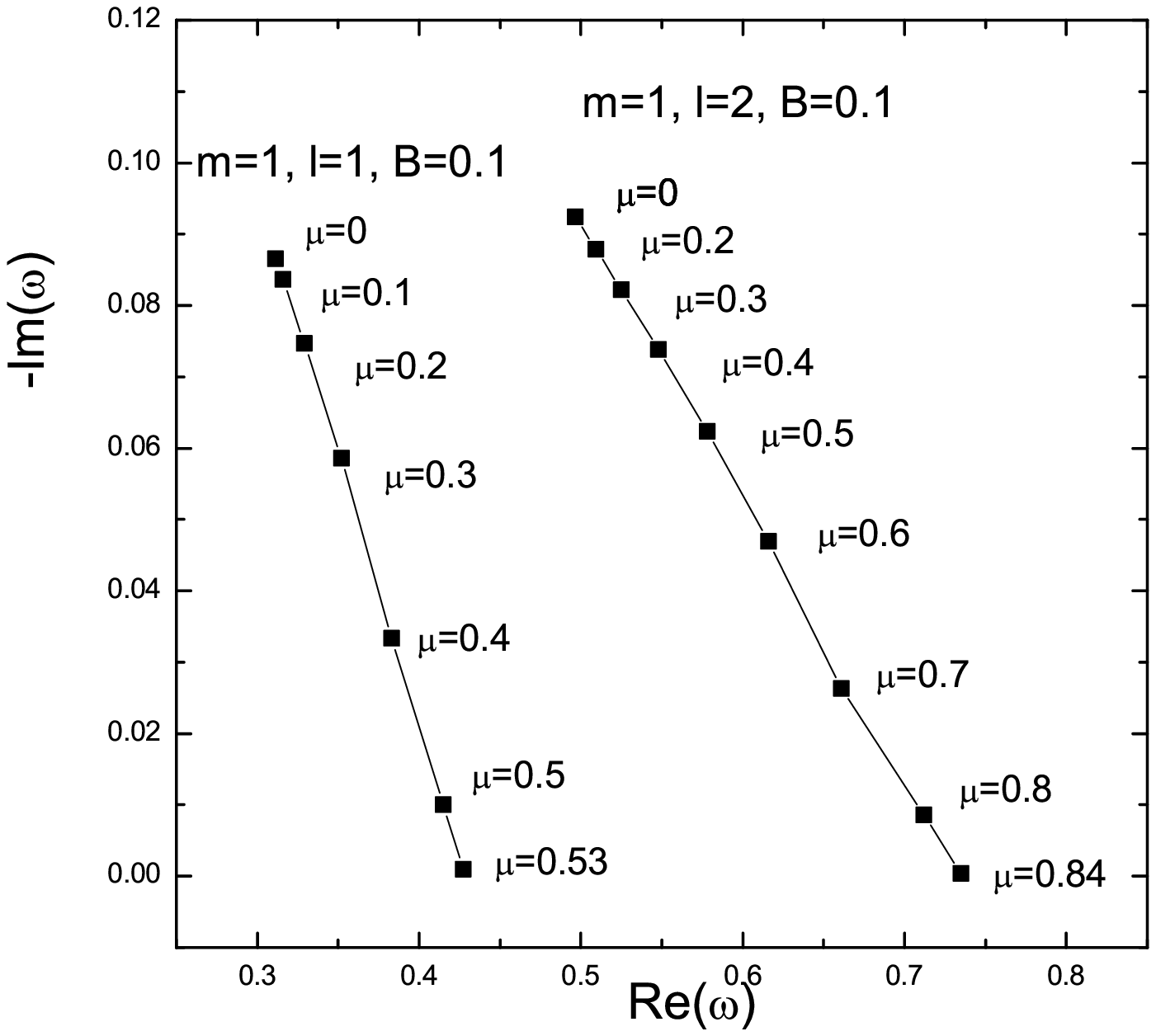}
\caption{ Fundamental quasinormal modes of the massive scalar field for Ernst black holes,  for magnetic field $B = 0.1, m =1, l = 1, 2$.
Given $l = 1$  for example, the  bigger the scalar field mass $\mu$ is, the  smaller the imaginary part of QNM becomes. When the threshold value of $\mu$ ($\mu = 0.53$) is exceeded, the particular QNMs disappear. }
\end{figure}


\begin{references}
\bibitem{black hole} V.P. Frolov and I.D. Novikov, Black Hole Physics: Basic Concept and New Developments (Kluwer Academic, Dordrecht, 1998).

\bibitem{LISA} E. Berti, V. Cordoso, and C.M. Will, Phys. Rev. D \textbf{73}, 064030 (2006).

\bibitem{hydro} D.T. Son and A.O. Starinets, Ann. Rev. Nucl. Part. Sci. \textbf{57}, 95 (2007).

\bibitem{review} K.D. Kokkotas and B.G. Schmidt, Living Rev. Rel. \textbf{2}, 2 (1999);
      H.P. Nollert, Class. Quantum Grav.  \textbf{16}, R159 (1999);
      E. Berti, V. Cardoso, and A.O. Starinets, Class. Quantum Grav.  \textbf{26}, 163001 (2009);
      R.A. Konoplya and A.V. Zhidenko, Rev. Mod. Phys. \textbf{83}, 793 (2011);

\bibitem{will} L.E. Simone and C.M. Will, Class. Quantum Grav.  \textbf{9}, 963 (1992).

\bibitem{Konoplya} R.A. Konoplya, Phys. Lett. B \textbf{550}, 117 (2002);
                  R.A. Konoplya and A.V. Zhidenko, Phys. Lett. B \textbf{609}, 377 (2005);

\bibitem{Koyama} H. Koyama and A. Tomimatsu, Phys. Rev. D \textbf{63}, 064032 (2001);
                 H. Koyama and A. Tomimatsu, Phys. Rev. D \textbf{64}, 044014 (2002).

\bibitem{Wang} L.H. Xue, B. Wang and R.K. Su, Phys. Rev. D \textbf{66}, 024032 (2002);



\bibitem{WKB} S. Iyer and C.M. Will, Phys. Rev. D \textbf{35}, 3621 (1987);
              B.F. Schutz and C.M. Will, Astrophys. J. Lett. \textbf{291}, L33 (1985).
              R.A. Konoplya, Phys. Rev. D \textbf{68}, 024018 (2003);

\bibitem{resonance} A. Ohashi and M.A. Sakagami, Class. Quantum Grav.  \textbf{21}, 3973 (2004).

\bibitem{Leaver} E.M. Leaver, Proc. R. Soc. A \textbf{402}, 285 (1995).

\bibitem{wuchen} C. Wu and Z. Ren, Phys. Rev. C \textbf{83}, 025805 (2011);
                R. Xu, C. Wu and Z. Ren, Int. J. Mod. Phys. E  \textbf{23}, 1450078 (2014).

\bibitem{SNZhang} W. M. Zhang, Y. Lu, and S. N. Zhang, arXiv:astro-ph/0501365.

\bibitem{Magnetized BH QNM} R.A. Konoplya and R.D.B. Fontana, Phys. Lett. B \textbf{659}, 375 (2008).

\bibitem{Ernst} F.J. Ernst, J. Math. Phys, \textbf{17}, 54 (1976).

\bibitem{Brito} R. Brito, V. Cardoso and P. Pani, Phys. Rev. D \textbf{89}, 104045 (2014);
\end{references}
\end{document}